\documentclass[pra,twocolumn,superscriptaddress,aps,showpacs]{revtex4-1}

\usepackage{hyperref}
\usepackage{graphicx}
\usepackage{amsmath}
\usepackage{amsfonts}
\usepackage{amssymb}
\usepackage{times}
\usepackage{subfigure}
\usepackage[usenames,dvipsnames]{color}
\usepackage{bm}
\everymath{\displaystyle}
\begin{document}
\title{Dynamics of phase separation in two species Bose-Einstein condensates 
       with vortices}

\author{Soumik Bandyopadhyay}
\affiliation{Physical Research Laboratory,
             Ahmedabad - 380009, Gujarat,
             India}
\affiliation{Indian Institute of Technology Gandhinagar,
             Gandhinagar - 382424 , Gujarat,
             India}

\author{Arko Roy}
\affiliation{Physical Research Laboratory,
             Ahmedabad - 380009, Gujarat,
             India}

\author{D. Angom}
\affiliation{Physical Research Laboratory,
             Ahmedabad - 380009, Gujarat,
             India}

\date{\today}


\begin{abstract}
We examine the dynamics associated with the miscibility-immiscibility 
transition of trapped two-component Bose-Einstein condensates (TBECs) of dilute
atomic gases in presence of vortices. In particular, we consider TBECs of Rb 
hyperfine states, and Rb-Cs mixture. There is an enhancement of the 
phase-separation when the vortex is present in both condensates. In the case of
a singly charged vortex in only one of the condensates, there is enhancement 
when the vortex is present in the species which occupy the edges at 
phase-separation. But, suppression occurs when the vortex is in the species 
which occupies the core region. To examine the role of the vortex, we quench 
the inter-species interactions to propel the TBEC from miscible to immiscible 
phase, and use the time dependent Gross-Pitaevskii equation to probe the 
phenomenon of phase-separation. We also examine the effect of higher charged 
vortex.
\end{abstract}



\maketitle

\section{Introduction}\label{Introduction}
Miscibility-immiscibility phase transition in a TBEC of dilute atomic gases, is 
a novel quantum phenomenon. It is also referred to as phase-separation, and 
provides a scheme to understand the physics governing a wide range of processes 
such as pattern formation, nonlinear excitations, dynamical and interface
instabilities~\cite{sas_09, kob_11, kad_12}. Further more, it is the key to
gain insights on phenomena such as quantum phase transition and criticality, 
symmetry breaking phenomena, Kibble-Zurek mechanism~\cite{sab_11}, collective
modes~\cite{tic_13} etc. In experiments, TBECs consisting of two different 
atomic species~\cite{mod_02, tha_08, mcc_11, ler_11, pas_13, wac_15, wan_16}, 
different isotopes of the same atomic species~\cite{pap_08, sug_11} or two
different hyperfine spin sates~\cite{mya_97, hal_98, toj_10} have been realized.
During the past two decades numerous theoretical studies have examined the 
static~\cite{ho_96,gau_11} and dynamical properties of 
phase-separation~\cite{nav_09, pat_13, lee_16}. From these studies it is clear 
that in the Thomas-Fermi (TF) limit at zero temperature the relative values of
the intra and inter-species interactions determine the miscibility or
immiscibility of the condensates. The condition for the phase-separation is the
inequality $g_{12} > \sqrt{g_{11}g_{22}} $, where $g_{12}$ is the inter-species
interaction strength, and $g_{kk}$ is the intra-species interaction of the
$k$th species. Based on this the TBEC can be driven from one phase to other by 
tuning the interaction strengths. However, an important point to be noted is 
that the derivation of the inequality assumes the TBEC to be in the ground 
state, that is, in absence of topological defects and impurities in the 
condensates. This aspect requires due investigation as there can be deviations 
from the inequality when vortices are present in the condensates. The effects 
of finite temperature on the dynamics of miscibility-immiscibility 
phase-separation of a TBEC is a topic of recent interest~\cite{lee_16}. In 
addition, suppression of phase-separation of a TBEC at finite temperatures has 
been reported~\cite{roy_15}. It has also been shown in theoretical 
investigations that inclusion of kinetic energy terms in the total energy 
expression of a TBEC, results in partial or complete suppression of 
phase-separation~\cite{wen_12}. This is to be contrasted with the TF 
approximation where the kinetic energy term is neglected.

In this work we theoretically investigate the effect of vortices on the 
dynamics of phase-separation in TBECs. An obvious way in which the vortices can 
influence the dynamics of phase-separation is through the centrifugal force 
arising from the associated superfluid flow. Thus, depending on the species in 
which vortex is introduced there can either be enhancement or suppression of 
phase-separation. In terms of experimental realizations, vortex in TBECs may be
produced using the method of phase imprinting~\cite{lea_02, kaw_04}, stirring 
of the condensates by Gauss-Laguerre laser beams~\cite{mar_97}, rotating the 
trapping potential~\cite{mar1_98, mar2_98}, through evaporative cooling 
process~\cite{marsh_99} or by interconversion between the two components of in 
the case of a TBEC with two hyperfine states~\cite{mat_99}. Other than the 
effects on the dynamics of phase separation, vortices in condensates are 
topological defects which are essential ingredient of several novel phenomena.
For the present work we examine the effects of when a vortex is present in one
of the condensate species in a TBEC, as well as vortices are present in both 
the species. In addition, we also investigate the effects of the charge of the
vortex, and it is expected that higher charged vortices shall 
have a larger effect. However, equally important is the dynamics and stability 
associated with a vortex with higher charge or vorticity. 

The paper is organized as follows. In Sec.~\ref{Theoretical_methods} we
formulate the dynamics of phase-separation of a TBEC at zero temperature, in
the Gross-Pitaeveskii framework, and discuss on the effects of centrifugal 
force associated with vortex induced superfluid flows in the condensates. 
Sec.~\ref{Numerical_methods} provides a brief description of the numerical 
schemes used to probe the phenomenon of phase-separation, and investigate on 
the dynamics associated with it. In Sec.~\ref{Results_and_discussions}, we 
present the results describing the vortex induced enhancement or suppression in 
miscibility-immiscibility transition of the TBECs depending on its presence
in the species. We also report the results from our further investigations on 
the dynamics in the presence of higher charged vortex. We conclude with the 
key highlights of our finding in Sec.~\ref{Conclusions}.


\section{Theoretical methods}\label{Theoretical_methods}
 
In mean field approximation, the time evolution of the order parameters of an
interacting, trapped TBEC system at $T = 0$K, are governed by a pair of coupled 
Gross-Pitaeveskii (GP) equations
 \begin{eqnarray}
  && \left[
     -\frac{\hbar^{2}}{2m_k}\nabla^{2} + V_{k}(\mathbf{r}) 
     + \sum_{j = 1}^{2} g_{kj}|\Psi_j(\mathbf{r}, t)|^2
     \right]
     \Psi_{k}(\mathbf{r}, t)
     \nonumber\\
  && = {\rm i}\hbar \frac{\partial \Psi_{k}(\mathbf{r}, t)}{\partial t},
 \label{eq_cgp}
 \end{eqnarray}
where, $k = 1, 2$ is species index, $\Psi_k$ is the condensate
wavefunction of the $k$th species, and $V_{k}(\mathbf{r})$ is the trapping
potential. The intra and inter-species interaction strengths are given by 
$g_{kk} = 4\pi\hbar^2a_{kk}/m_k$, and $g_{kj} = 2\pi\hbar^2a_{kj}/m_{kj}$, 
respectively. Here, $a_{kk}$ and $a_{kj}$ are the intra and inter-species 
$s$-wave scattering lengths of atoms, $m_{k}$ is mass of the $k$th species, and
$m_{kj} = m_{k}m_{j}/(m_{k}+m_{j})$ is the reduced mass. The order parameters 
or wave functions of each of the species are normalized to the total number of 
atoms in the condensates
 \begin{equation}
  N_{k} = \int{\rm d}\mathbf{r}|\Psi_{k}(\mathbf{r})|^{2}.
 \end{equation}
With these considerations and definitions, the total energy of the TBEC system 
is
 \begin{eqnarray}
   E = && \int{\rm d}\mathbf{r}\hspace{0.05cm}
          \biggr[
          \sum_{k = 1}^{2}
          \biggr(
          \frac{\hbar^{2}}{2m_{k}}|\nabla\Psi_{k}(\mathbf{r})|^{2} 
          + V_{k} (\mathbf{r})|\Psi_{k}(\mathbf{r})|^{2}
          \nonumber\\
       && \hspace{1cm} + \frac{g_{kk}}{2}|\Psi_{k}(\mathbf{r})|^{4}
          \biggr)
          + g_{12}|\Psi_{1}(\mathbf{r})|^{2}|\Psi_{2}(\mathbf{r})|^{2}
          \biggr],
 \label{total_energy}
 \end{eqnarray}
where $V_k$ is taken to be harmonic oscillator potential which is of the
form
 \begin{equation}
   V_{k}(\mathbf{r}) = V_{k}(x,y,z) = \frac{1}{2}m_{k}\omega_{k}^{2}(x^{2}
   + \alpha_{k}^{2}y^{2} + \lambda_{k}^{2}z^{2}).
 \label{trap_3d}
 \end{equation}
Here, $\omega_{k}$ is frequency of the trap along $x$ direction,
$\alpha_{k}$, $\lambda_{k}$ are the anisotropy parameters. For the present
study, we consider the atoms of both species to be trapped in the same
potential, that is, $\omega_{1} = \omega_{2} = \omega_x$,
$\alpha_{1} = \alpha_{2} = \alpha = \omega_y/\omega_x$,
and $\lambda_{1} = \lambda_{2} = \lambda = \omega_z/\omega_x$.
Furthermore, we define the oscillator length to be
$a_{\rm osc} = \sqrt{\hbar/(m_{1}\omega_x)}$, and  energy quanta
$\hbar\omega_x$ which correspond to convenient length and energy scale
of the system. To render the coupled GP equations in dimensionless form, we 
scale the co-ordinates to $\tilde{x} = x/a_{\rm osc}$, 
$\tilde{y} = y/a_{\rm osc}$, $\tilde{z} = z/a_{\rm osc}$, time to 
$\tilde{t} = t\omega_x$, and total energy to $\tilde{E} = E/(\hbar\omega_x)$. 
The order parameters then follow the transformations
 \begin{equation}
   \Phi_{k}(\tilde{x},\tilde{y},\tilde{z}) =
   \sqrt{\frac{a_{\rm osc}^{3}}{N_{k}}}\Psi_{k}(x,y,z).
 \label{sd_odr_pmtr}
 \end{equation}
Defining $m_{\rm r} = m_{1}/m_{2}$, the total energy in 
Eq.~(\ref{total_energy}) in dimensionless form is
 \begin{eqnarray}
   \tilde{E} = &&   
                  \int d\tilde{x}d\tilde{y} d\tilde{z}
                  \biggr \{
                  \frac{N_{1}}{2}\biggr [ |\nabla\Phi_{1}|^{2} 
                  + (\tilde{x}^{2} + \alpha^{2}\tilde{y}^{2} + 
                  \lambda^{2}\tilde{z}^{2})|\Phi_{1}|^2 
                  \nonumber\\
               && + N_{1}\tilde{g}_{11}|\Phi_{1}|^{4}\biggr ] 
                  + \frac{N_{2}}{2}\biggr [m_{\rm r}|\nabla\Phi_{2}|^{2}
                  + \frac{1}{m_{\rm r}}(\tilde{x}^{2} + \alpha^{2}\tilde{y}^{2}
                  \nonumber\\
               && + \lambda^{2}\tilde{z}^{2})|\Phi_{2}|^2 
                  + N_{2}\tilde{g}_{22}|\Phi_{2}|^{4}\biggr ] 
                  + N_{1}N_{2}\tilde{g}_{12}|\phi_{1}|^{2}|\Phi_{2}|^{2}
                  \biggr \}
                  \nonumber\\
 \label{scld_energy}
 \end{eqnarray}
where, $\tilde{g}_{11} = 4 \pi a_{11}/a_{\rm osc}$, 
$\tilde{g}_{22} = m_{\rm r} 4 \pi a_{22}/a_{\rm osc}$ and
$\tilde{g}_{12} = 2\pi (m_{1} + m_{2})a_{12}/m_{2}a_{\rm osc}$. For notational
convenience, here after we drop the tilde from the transformed quantities. The 
scaled coupled GP equations can then be expressed as
 \begin{eqnarray}
   && \biggr[
      - \frac{1}{2}\nabla^{2}
      + \frac{1}{2}(x^{2} + \alpha^{2}y^{2}+\lambda^{2}z^{2})
      \nonumber\\
   && + \sum_{j = 1}^{2}G_{1j}|\Phi_{j}(x, y, z, t)|^{2}
      \biggr]
      \Phi_{1}(x, y, z, t)
      = {\rm i}\frac{\partial\Phi_{1}(x, y, z, t)}{\partial t},
      \nonumber\\
   && {\rm  and}
      \nonumber\\
   && \biggr[
      - \frac{m_{\rm r}}{2}\nabla^{2}
      + \frac{1}{2m_{\rm r}}(x^{2} + \alpha^{2}y^{2}+\lambda^{2}z^{2})
      \nonumber\\
   && + \sum_{j = 1}^{2}G_{2j}|\Phi_{j}(x, y, z, t)|^{2}
      \biggr]
      \Phi_{2}(x, y, z, t)
      = {\rm i}\frac{\partial\Phi_{2}(x, y, z, t)}{\partial t},
      \nonumber\\
 \label{scld_cgp}
 \end{eqnarray}
where, $g_{11} = N_{1}4\pi a_{11}/a_{\rm osc}$, 
$g_{22} = m_{\rm r} N_{2} 4\pi a_{22}/a_{\rm osc}$ and
$g_{kj} = N_{j} 2\pi(m_{1} + m_{2}) a_{kj}/m_{2} a_{\rm osc}$.
The TBEC system in our study is confined in a quasi-two-dimensional (quasi-2D)
harmonic trap. This is achieved by considering the axial frequency of the trap,
$\omega_{z}$, to be much larger than the frequencies along $x$ and $y$
directions, therefore, $\lambda\gg 1$, and to maintain radial symmetry we take
$\alpha = 1$. This condition allows us to factorize the order parameters in the 
following form
 \begin{equation}
     \Phi_{k}(x, y, z, t) = \psi_{k}(x, y, t)\chi_{k}(z),
     \label{fact_odr_pm}
 \end{equation}
where, $\chi_{k}(z)$ are normalized ground states of the condensates along the 
axial direction. Substituting Eqns.~(\ref{fact_odr_pm}) in 
Eqns.~(\ref{scld_cgp}), and then integrating over $\chi_{k}(z)$, we obtain the 
following scaled coupled GP equations in 2D
 \begin{eqnarray}
  && \biggr[
     - \frac{1}{2}\nabla_{\perp}^{2} + \frac{1}{2}(x^{2} + \alpha^{2}y^{2}) 
     + \sum_{j = 1}^{2}\mathcal{G}_{1j}|\psi_{j}(x, y, t)|^{2}
     \biggr]
     \nonumber\\
  && \times\psi_{1}(x, y, t) 
     = {\rm i}\frac{\partial\psi_{1}(x, y, t)}{\partial t},
     \nonumber\\
  && {\rm and}
     \nonumber\\
  && \biggr[
     - \frac{m_{\rm r}}{2}\nabla_{\perp}^{2} 
     + \frac{1}{2m_{\rm r}}(x^{2} + \alpha^{2}y^{2}) 
     + \sum_{j = 1}^{2}\mathcal{G}_{2j}|\psi_{j}(x, y, t)|^{2}
     \biggr]
     \nonumber\\
  && \times\psi_{2}(x, y, t) 
     = {\rm i}\frac{\partial\psi_{2}(x, y, t)}{\partial t}, 
 \label{time_dep_gp_2d}
 \end{eqnarray}
where, $\nabla_{\perp}^2 = \partial_x^2 + \partial_y^2$,
$\mathcal{G}_{11} = 2N_{1}\sqrt{2\pi\lambda}a_{11}/a_{\rm osc}$,
$\mathcal{G}_{22} = m_{\rm r} 2N_{2}\sqrt{2\pi\lambda}a_{22}/a_{\rm osc}$ and
$\mathcal{G}_{kj} = N_{j}(m_{1} + m_{2})\sqrt{2\pi\lambda}a_{kj}/m_{2}
a_{\rm osc}$. With these definitions the time independent coupled GP equations 
are
 \begin{eqnarray}
  && \biggr[ 
     - \frac{1}{2}\nabla_{\perp}^{2} +  \frac{1}{2}(x^{2} + \alpha^{2}y^{2})
     + \sum_{j = 1}^{2}\mathcal{G}_{1j}|\psi_{j}(x, y)|^{2}
     \biggr]
     \psi_{1}(x, y)
     \nonumber\\
  && = \mu_{1}\psi_{1}(x, y),
     \nonumber\\
  && {\rm and}
     \nonumber\\
  && \biggr[
     - \frac{m_{\rm r}}{2}\nabla_{\perp}^{2} 
     + \frac{1}{2m_{\rm r}}(x^{2} + \alpha^{2}y^{2}) 
     + \sum_{j = 1}^{2}\mathcal{G}_{2j}|\psi_{j}(x, y)|^{2}
     \biggr]
     \nonumber\\
  && \times\psi_{2}(x, y) = \mu_{2}\psi_{2}(x, y), 
 \label{time_indep_gp_2d}
 \end{eqnarray}
where, $\mu_k$ is the chemical potential of the $k$th species condensate.


\subsection{Phase-separation}
In the Thomas-Fermi (TF) limit~\cite{bay_96, ho_96}, depending on interaction 
strengths, the system can exhibit two distinct phases, miscible or immiscible
(phase-separated)~\cite{tri_00}. In the miscible phase, the condensates overlap 
with each other; whereas, they get spatially separated in immiscible phase. 
A measure to characterize these phases is the overlap integral~\cite{jai_11}
 \begin{equation}
 \Lambda =
        \frac{
        \biggr[
        \displaystyle\int\hspace{-0.2cm}\displaystyle\int{\rm d}x
        \hspace{0.05cm}{\rm d}y\hspace{0.05cm}n_{1}(x, y)
        \hspace{0.05cm}n_{2}(x,y)
        \biggr]^{2}
        }
        {
        \biggr[
        \displaystyle\int\hspace{-0.2cm}\displaystyle\int{\rm d}x
        \hspace{0.05cm}{\rm d}y\hspace{0.05cm}n_{1}^{2}(x, y)
        \biggr]
        \biggr[
        \displaystyle\int\hspace{-0.2cm}\displaystyle\int{\rm d}x
        \hspace{0.05cm}{\rm d}y\hspace{0.05cm} n_{2}^{2}(x, y)
        \biggr] 
        },
  \label{ovlp_measure}
 \end{equation}
where, $n_k(x, y) = |\psi_{k}(x, y)|^{2}$ is the density of the 
$k$th condensate species. A value of $\Lambda = 1$ implies complete overlap
between the condensates or the two species are completely miscible, and
complete phase-separation corresponds to $\Lambda = 0$. The criterion for 
phase separation, based on the Thomas-Fermi approximation and minimization of 
the  total energy  given in Eq.~(\ref{total_energy}), is 
$g_{12} > \sqrt{g_{11}g_{22}}$. It should, however, be mentioned that this 
condition is valid only at zero temperature, and in the absence of any 
topological defects. There are deviations from this criterion at $T\neq 0$ 
due to the presence of thermal atoms~\cite{roy_15}. In addition, the superflows 
associated with vortices in TBECs are expected to influence this criterion.


\subsection{Effect of vortices}
Employing Madelung transformation to the order parameter 
$\Psi_{k}(\mathbf{r}, t)$, we can express super fluid velocity as 
$\mathbf{v}_{k} = \hbar\nabla\theta_{k}/m_{k}$, where, 
$\theta_{k}(\mathbf{r}, t)$ is the phase of the order parameter. Then, the 
presence of a vortex in the condensate results in an additional superfluid flow
(super-flow), and around it the phase of the order parameter changes by 
$2\pi l$, where $l = \pm 1, \pm 2, \pm 3 ...$ $l$ is the charge or vorticity of
the vortex. Considering the vortex induced super-flow as purely azimuthal, the
velocity of the flow at a distance $R$ from the vortex core 
is~\cite{pet_bk_08, ued_bk_10}
\begin{equation} 
  \mathbf{v}_{k}(R) = \frac{l\hbar}{m_{k}R}\mathbf{e}_\phi .
\end{equation}
As a consequence of this
superflow, the atoms in the condensate experience a radially outward 
centrifugal force of magnitude 
\begin{equation}
  \mathbf{F}_{k}(R) = \frac{l^{2}\hbar^{2}}{m_{k}R^{3}}\mathbf{e}_R.
\end{equation}
>From this expression, it is evident that lower atomic mass is associated with 
stronger force; and the quadratic dependence on $l$ implies that the force is 
independent of the sign of vortex charge. Due to the centrifugal force 
the onset of phase-separation can be enhanced when the vortex is associated 
with the species that lies at the periphery at phase-separation. And, 
suppression when the vortex is associated with the species occupying the 
core at phase-separation. Thus, as mentioned earlier, the presence of vortex
modifies the criterion for phase-separation. This is investigated in more
detail or in a quantitative way numerically.

 The stability of the vortex is dependent on the vortex charge. In a quasi-2D 
single species condensates, a singly charged vortex is dynamically stable, and 
precesses on an equidensity circular contour. But, a vortex of charge greater 
than unity is unstable, and spontaneously decays into multiple singly charged 
vortices during evolution, even in absence of dissipation and external 
perturbations~\cite{pu_99, shi_04, kaw_04}. In the case of a TBEC, in the 
immiscible domain, the vortex core in condensate of one of the species is filled 
by the condensate atoms of the other species~\cite{mat_99}, and the vortex is 
considered as coreless. Then, the superflow around the vortex in one condensate 
influences the other species, which results in an additional interaction among 
the condensates, and is responsible for a range of dynamical phenomena in 
TBECs~\cite{rip_00, ohb_02}. However, stability of a higher charge vortex 
during its evolution, is now dependent on the miscibility or immiscibility of 
the condensates together with its presence in the condensates of the species. 
In the TF-limit,  the core size of a charge $l$ vortex is
\begin{equation} 
  \xi_{k} = \frac{l}{\sqrt{2n_{k}(\mathcal{G}_{kk} + \mathcal{G}_{kj})}},
\end{equation}
where, $n_{k}$ is taken to be the local TF density of the condensate at the
trap center in absence of the vortex~\cite{ish_13, gau_13}. Considering the
larger core size and centrifugal force with higher $l$, the enhancement or
suppression of phase-seperation with a vortex in one of the species is 
more pronounced with higher $l$. However, the dynamics of the 
miscible-immiscible transition would exhibit complex patterns as vortex with
$l>1$ decays to vortices with unit charge.


\section{Numerical methods}\label{Numerical_methods}

The first step of the computations is to obtain the equilibrium solution in 
the miscible domain as the initial state. For this we numerically solve 
Eqns.~(\ref{time_indep_gp_2d}) in imaginary time using the split-step 
Crank-Nicholson method adapted for binary condensates. Furthermore, we use the 
numerical procedure of phase-imprinting technique to introduce a vortex of 
charge $l$ by taking~\cite{dob_99, lea_02} the order parameter as
 \begin{equation}
  \psi_{k}(x, y) = |\psi_{k}(x, y)|\exp
  \left[{\rm i}l\tan^{-1}\left(\frac{y-y_{0}}{x-x_{0}}\right)\right],
 \end{equation}
where $(x_{0}, y_{0})$ is the location of the vortex in the
condensates. To study the dynamics of phase-separation, we consider the 
equilibrium state solution obtained from the imaginary time propagation, and 
then evolve it over real time. For the present purpose, the 
inter-species scattering length $a_{12}$ is adiabatically quenched from a value
corresponding to the miscible phase of the TBEC to a value satisfying the 
phase-separation condition. The tuning of $a_{12}$ is experimentally possible 
through magnetic Feshbach resonance. We investigate on the dynamics of the
considered TBECs during this quench in absence and presence of a vortex in the
condensates and then evolve them freely for $750 {\rm ms}$ to examine post 
quench dynamics of the systems.
\begin{figure}[ht]
    \includegraphics[height = 5.5cm]{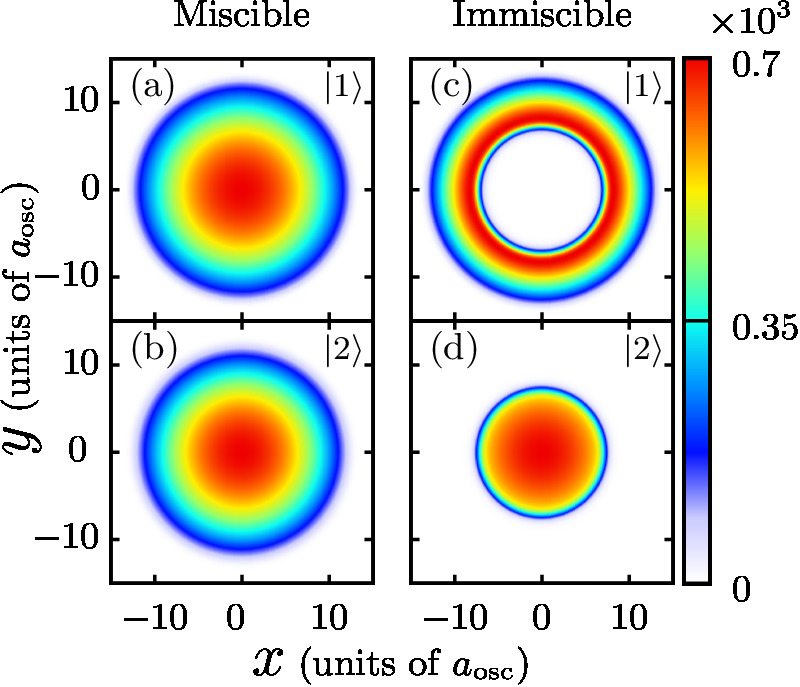}
    \caption{(Color online)
            (a) and (b) show the density profiles of the Rb condensates in the
            two different hyperfine states when the TBEC is in miscible phase 
            at $a_{12} = 70a_{0}$. (c) and (d) show the density profiles in the 
            immiscible phase at $a_{12} = 100a_{0}$. In this phase, 
            condensates of the species $|1\rangle$ and $|2\rangle$ become 
            shell-condensate and core-condensate respectively. The color bar 
            represents number density of atoms in the condensates in units of
            $a_{\rm osc}^{-2}$, where the considered oscillator length 
            $a_{\rm osc} = 1.9\mu$m.
            }
    \label{mis_immis}
\end{figure}


\section{Results and discussions}\label{Results_and_discussions}
As a representative example to study the dynamics of 
phase-separation in the presence of a vortex, we first consider the BEC
mixture in the hyperfine states $|F = 1, m_f = -1\rangle \equiv |1\rangle$
and $|F = 2, m_f = +1\rangle \equiv |2\rangle$ of $^{87}$Rb, which has
been experimentally obtained to probe different static and dynamic properties
of a TBEC~\cite{hal_98, mer_07}. For this mixture $m_{\rm r} = 1$ as 
$m_1 = m_2$. Following the experimental realization~\cite{mer_07} we
consider a rotationally symmetric harmonic trap with 
$\omega_x = \omega_y = 2\pi\times 30.832\hspace{0.1cm}$Hz. And, to satisfy
quasi-2D condition we consider $\omega_z = 100.0 \omega_x$ so that 
$\mu_{k} \ll \hbar\omega_{z}$, and take equal total number of atoms in the 
condensates as $N_1 = N_2 = 10^5$. The intra-species scattering lengths,
$a_{11}$ and $a_{22} $, are $100.4a_{0}$ and $95.44a_{0}$~\cite{ego_13},
respectively, where $a_0$ is Bohr radius. For these values, the TBEC is in
the immiscible domain when $a_{12} \geqslant 97.9a_0$. To steer the 
TBEC from the miscible to immiscible phase, we tune $a_{12}$ from $70a_{0}$ 
to $100a_{0}$. As mentioned earlier, this is possible through the
magnetic Feshbach resonance~\cite{mar_02, erh_04, toj_10}. In the immiscible
phase, the energetically favorable solution at equilibrium is a 
shell-structured geometry, in which the atoms having smaller scattering length
in $|2\rangle$ state, occupy the central region of the trap, here after 
referred to as the {\em core-condensate}. And, the atoms with the larger 
scattering length in $|1\rangle$ state form a lower density shell about 
the core-condensate, thus, referred to as {\em shell-condensate}. The 
density profiles of the core and shell-condensate for $a_{12} = 100 a_0$ 
are shown in Fig.~\ref{mis_immis}(c) and (d) .

As an example of TBEC with unequal masses we consider $^{87}$Rb-$^{133}$Cs
TBEC~\cite{mcc_11}, referred to as Rb-Cs TBEC for compact notation, for this mixture $m_{\rm r} \approx 0.65$. The results for other TBECs like 
$^{87}$Rb-$^{39}$K, $^{87}$Rb-$^{23}$Na, etc are expected to be 
qualitatively similar. For convenience, we label $^{87}$Rb and $^{133}$Cs to 
be the first and second species, respectively, and take $N_1 = N_2 = 10^4$. We 
consider this mixture in a rotationally symmetric trap with
$\omega_x=\omega_y = 2\pi\times 8\hspace{0.1cm}$Hz, and
$\omega_z = 40.0 \omega_x$ so that $\mu_{k} \ll \hbar\omega_{z}$. The 
intra-species scattering lengths of the $^{87}$Rb and $^{133}$Cs atoms, 
$a_{11}$ and $a_{22} $, are $99a_{0}$ \cite{mar_02} and $280a_{0}$ 
\cite{chi_04}, respectively.  Hence, the phase-separation condition is  
$a_{12} \geqslant 162.8a_0$. We drive the Rb-Cs TBEC from from the 
miscible to immiscible phase by varying $a_{12}$ from $50a_{0}$ to $175a_{0}$ 
which is possible through magnetic Feshbach resonance~\cite{pil_09}. In the 
immiscible phase the ground state density distribution of the system has  
shell-structured geometry like in the previous system. However, despite of 
inter-species scattering length of Cs is much larger than that of Rb, In the 
immiscible phase the heavier Cs atoms occupy the central region of the trap or 
form the core-condensate, and the lighter Rb atoms are at the edge or form the 
shell-condensate. This is despite the much larger intra-species scattering 
length of Cs atoms as this configuration tends to minimize the total energy
by lowering the contribution from the trapping potential. 
 \begin{figure}[ht]
    \includegraphics[height = 4.0cm]{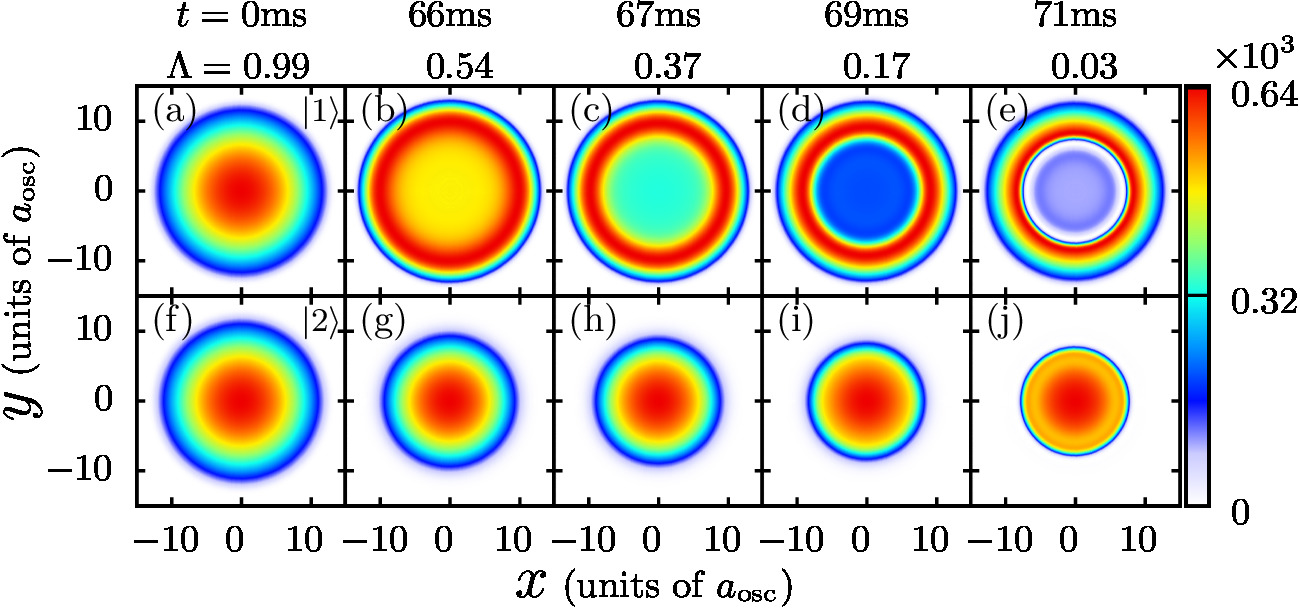}
    \caption{(Color online)
            Shows the time evolution of density profiles of the Rb-condensates
            in $|1\rangle$ and $|2\rangle$ states, during the quench of 
            $a_{12}$, in absence of vortices in the condensates. The first
            and second row show the density profiles of the condensates in
            $|1\rangle$ and $|2\rangle$ states respectively, at $0$ms, $66$ms,
            $67$ms $69$ms and $71$ms. At these instants, the values of $a_{12}$
            are $70a_{0}$, $96.7a_{0}$, $97.3a_{0}$ $98.0a_{0}$ and $98.7a_{0}$
            respectively. Values of the overlap measure $\Lambda$ at 
            corresponding time instants are mentioned at the top of each 
            column. The color bars represent number density of atoms in the 
            condensates in units of $a_{\rm osc}^{-2}$ where 
            $a_{\rm osc} = 1.9\mu$m.
            }
    \label{wovrtx}
 \end{figure}


\subsection{Dynamics of phase-separation without vortex}
At initial time, the equilibrium state solution of the TBEC in Rb-hyperfine
states is obtained in miscible phase by considering $a_{12} = 70a_{0}$, and  
the corresponding density profiles of the condensates are shown in 
Fig.~\ref{wovrtx}(a) and (f). The condensates then have maximal overlap, and
hence $\Lambda = 0.99$. Now, we increase $a_{12}$ at the rate of 
$0.41\hspace{0.1cm}a_{0}/{\rm ms}$~\cite{mer_07}. The evolution of the 
condensate density profiles during the quench are shown in Fig.~\ref{wovrtx},
and there is an increase in the total energy of the TBEC as the interaction
energy increases. However, after phase-separation, when $a_{12} > 97.9a_0$, the
overlap between the condensates becomes negligible, and therefore, the 
contribution to the total energy from the inter-species interaction is 
negligible. On the other hand, the higher $a_{12}$ enhances the gradient of the 
density profiles at the interface, and as a consequence, the kinetic energies 
of the condensates are increased. This in turn enhances the total energy. In 
this phase, the condensate of the $|1\rangle$ species surrounds the condensate 
of the $|2\rangle$ species in shell geometry. As example, the density profiles 
of the condensates at $71{\rm ms}$, with a corresponding value of 
$a_{12} =98.7a_{0}$, are shown in Fig.~\ref{wovrtx}(e) and (j). From the 
figures, it is evident that the TBEC is in immiscible phase, and 
$\Lambda = 0.04$. We, therefore, stop quench after $a_{12}$ attains the value 
of $100a_{0}$ at $74$ms. We then observe the free evolution of the density 
profiles. At later times, the condensates continue to be in this geometry while
exhibiting oscillations in the overlap with frequency 
$\nu \approx 185\hspace{0.1cm}$Hz, which is larger than the radial trap 
frequency $\nu_{x} = 30.832\hspace{0.1cm}$Hz.

In a similar way, we obtain the initial equilibrium solution for the Rb-Cs 
TBEC in the miscible by considering $a_{12} = 50a_{0}$, and has $\Lambda=1.0$. 
We then quench $a_{12}$ by increasing at the rate of 
$1.58\hspace{0.1cm}a_{0}/{\rm ms}$~\cite{mcc_11, ler_11}. The adiabaticity of 
the quench is verified by obtaining the stationary ground state solutions of 
the TBEC at the intermediate values of $a_{12}$. As in the previous case, the 
total energy of the TBEC increases with the increase of $a_{12}$, and the time 
evolution of the density profiles of the condensates 
are qualitatively similar. After phase-separation, when $a_{12} > 162.8a_0$,
as mentioned earlier the Rb-condensate surrounds the Cs-condensate in a shell 
geometry. In this geometry, the enhanced inter-species interaction makes the 
size of the pancake-shaped Cs-condensate smaller than its size in the miscible 
phase. This reduces the trapping potential energy of the Cs-condensate; but, 
the enhanced density increases the interaction energy of the Cs-condensate. 
The quench is stopped at $79{\rm ms}$ when $a_{12} = 175a_{0}$, and the 
overlap between the condensates has $\Lambda = 0.02$. We, then, observe the 
free evolution of the density profiles. At later times, the condensates 
continue to be in this geometry with an oscillation in the overlap at a 
frequency of $ \nu \approx 80\hspace{0.1cm}$Hz. Like in the previous case, this
is larger than the radial trap frequency $\nu_{x} = 8\hspace{0.1cm}$Hz. 
 \begin{figure}[ht]
    \includegraphics[height=4.0cm]{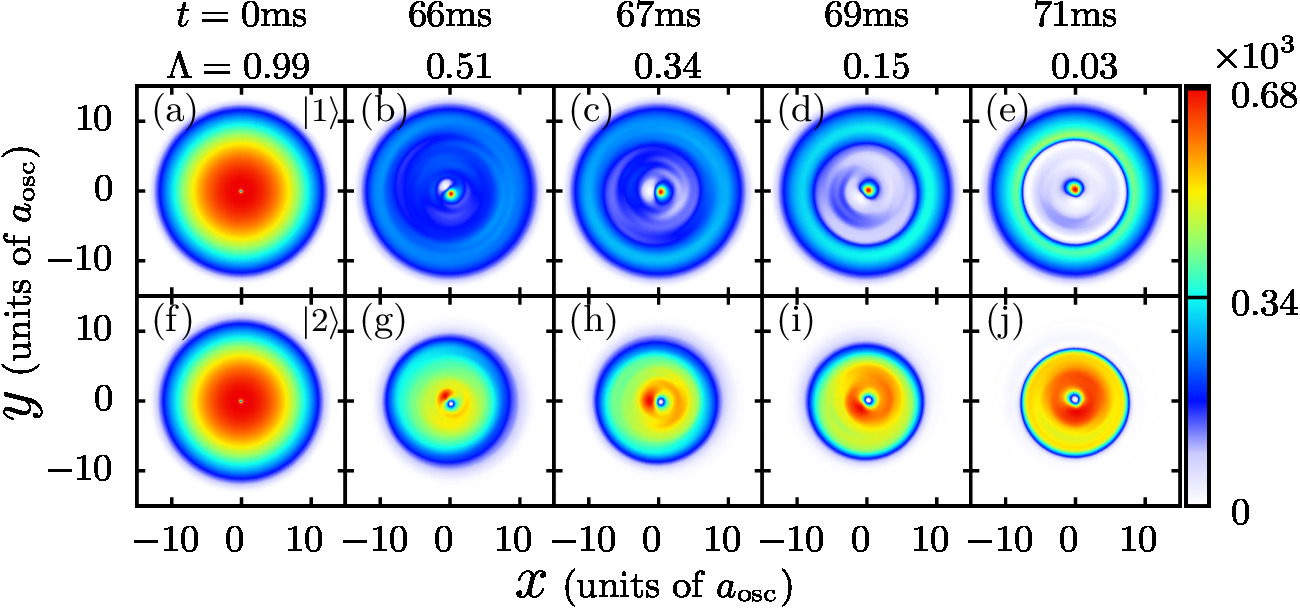}
    \caption{(Color online)
            Shows the time evolution of density profiles of the Rb-condensates
            in $|1\rangle$ and $|2\rangle$ states, during the quench of
            $a_{12}$, in presence of a singly charged vortex at the center of
            both condensates. The first and second row show the density 
            profiles of the condensates in $|1\rangle$ and $|2\rangle$ states 
            respectively, at $0$ms, $66$ms, $67$ms $69$ms and $71$ms. At these
            instants, the values of $a_{12}$ are $70a_{0}$, $96.7a_{0}$, 
            $97.3a_{0}$ $98.0a_{0}$ and $98.7a_{0}$ respectively. Values of 
            the overlap measure $\Lambda$ at corresponding time instants are 
            mentioned at the top of each column. The color bars represent 
            number density of atoms in the condensates in units of
            $a_{\rm osc}^{-2}$ where $a_{\rm osc} = 1.9\mu$m.
            }
    \label{wvrtx_q1}
 \end{figure}


\subsection{Presence of singly-charged vortex}


\subsubsection{Vortices in both the condensates}
\begin{figure}[ht]
    \includegraphics[height=6.0cm]{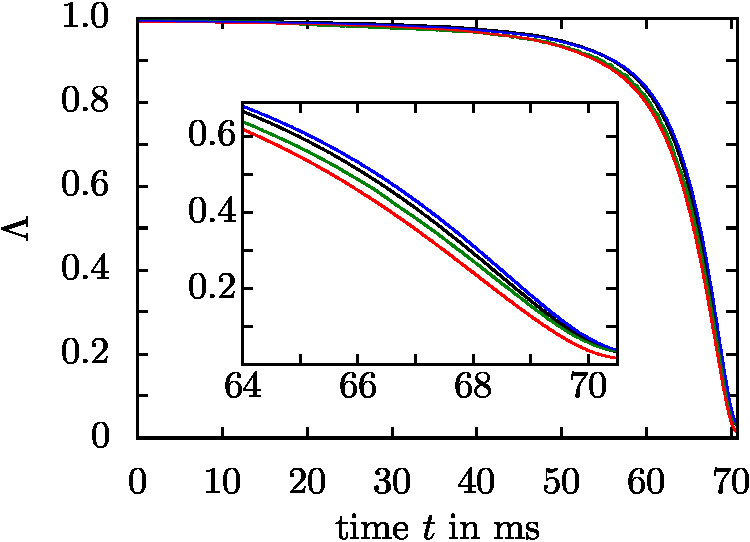}
    \caption{(Color online) 
            Shows the variation of the overlap measure $\Lambda$ with time
            as $a_{12}$ of the TBEC in Rb-hyperfine states quenched from 
            $70a_{0}$ to $100a_{0}$. The black curve shows this variation in
            absence of defects in the condensates, the green curve shows it
            when a singly charged vortex is present at the center of both 
            condensates; whereas, the red curve shows the variation when the 
            vortex is present only in the condensate of species $|1\rangle$ 
            (shell-condensate) but not in the condensate of species 
            $|2\rangle$ (core-condensate), and the blue curve shows the
            variation when the vortex is present only in the species 
            $|2\rangle$ but not in the condensate of species $|1\rangle$.
            }
 \label{lam_vs_time_q1}
\end{figure}

To examine the dynamics of the phase separation in the presence of a vortex in
the Rb hyperfine TBEC, we consider the equilibrium state with the same set of
parameters as previous. But, now we imprint singly charged vortices at the 
center of both the species. In experiments, this may be achieved by employing
topological phase imprinting techniques~\cite{lea_02}.  After obtaining the
equilibrium solution, like in the previous case, we quench $a_{12}$, to induce
miscibility-immiscibility phase transition in the system. During the course of
the evolution the vortices are displaced from the center and start to precess,
and the density profiles are as shown in Fig.~\ref{wvrtx_q1}. During the quench
there is an enhancement of the miscible-immiscible transition, which is evident
from the trend in the value of $\Lambda$ as shown in Fig.~\ref{lam_vs_time_q1}. From the figure there is a manifest faster decrease in $\Lambda$ when vortices 
are present in both the species. 

For the Rb-Cs as well we follow the same protocol of imprinting vortices in 
both the species, and quenching $a_{12}$ at the same rate as it was done
when the vortex was absent. Among the two condensates, due to the shorter 
healing length, the vortex core size in the Cs-condensate is smaller than in 
Rb. Here, the shorter healing length of Cs is on account of its larger mass and
scattering length. Unlike in the case of the Rb-hyperfine TBEC, the vortices in
the Rb-Cs TBEC remain at the center and the core size of the vortex in 
Rb-condensate increases. Following the values of $\Lambda$ during time 
evolution, as shown in Fig.~\ref{lam_vs_time_q1_rb_cs}, it is evident that 
there is an enhancement in the miscible-immiscible transition. To investigate 
further we imprint vortices with opposite charges, and find that the trend in 
the miscible-immiscible transition is independent of the sign of the vortex 
charges. In other words, it is the presence of the superflow which influences 
the onset of the phase-separation, but the direction of the superflow does not
impact on the transition. As evident from the comparison of the trends in 
Fig.~\ref{lam_vs_time_q1} and Fig.~\ref{lam_vs_time_q1_rb_cs}, the effect of 
the vortices is more pronounced in the case of Rb-Cs. This is on account of the 
difference in the masses and relative intra-species scattering lengths.
\begin{figure}[ht]
    \includegraphics[height=4.0cm]{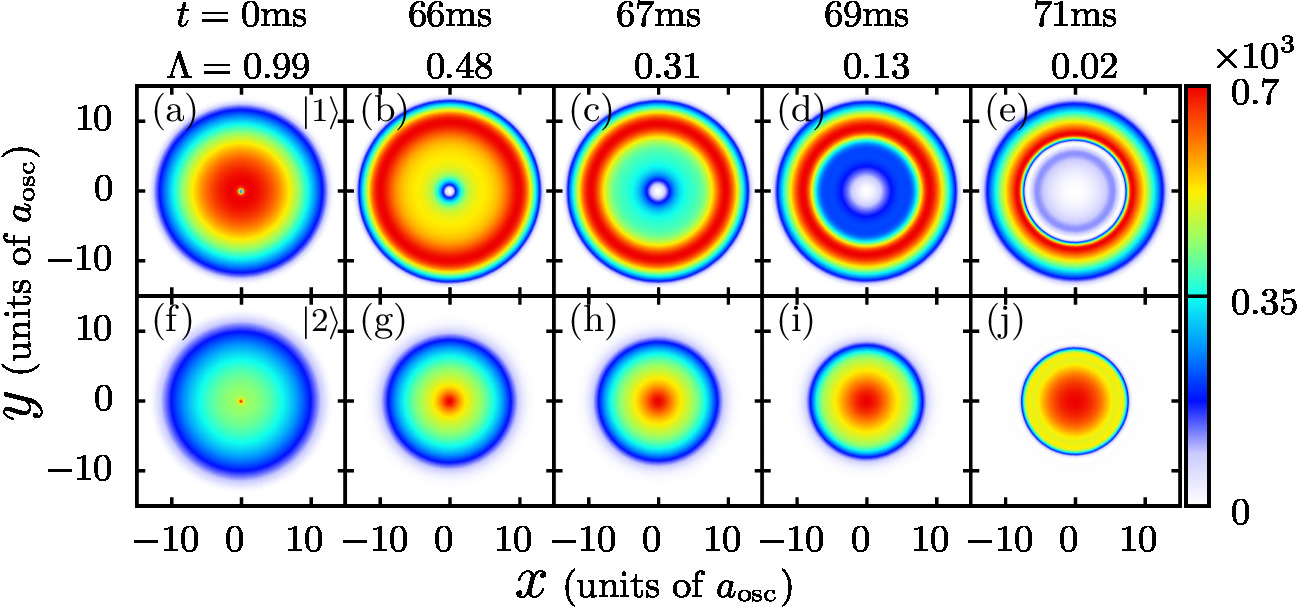}
    \caption{(Color online)
            Shows the time evolution of density profiles of the Rb-condensates
            in $|1\rangle$ and $|2\rangle$ states, during the quench of
            $a_{12}$, in presence of a singly charged vortex at the center of
            the shell-condensate. The first and second row show the density
            profiles of the condensates in $|1\rangle$ and $|2\rangle$ states
            respectively, at $0$ms, $66$ms, $67$ms $69$ms and $71$ms. At these
            instants, the values of $a_{12}$ are $70a_{0}$, $96.7a_{0}$,
            $97.3a_{0}$ $98.0a_{0}$ and $98.7a_{0}$ respectively. Values of
            the overlap measure $\Lambda$ at corresponding time instants are
            mentioned at the top of each column. The color bars represent
            number density of atoms in the condensates in units of
            $a_{\rm osc}^{-2}$ where $a_{\rm osc} = 1.9\mu$m.
            }
 \label{wvrtx_q1_cp1}
\end{figure}


\subsubsection{Vortex in shell-condensate}
To study the miscible-immiscible transition when vortex is 
present in only one of the species in  Rb-hypefine TBEC, we first 
examine the evolution of the TBEC with a vortex present only in the 
condensate of species $|1\rangle$. Like in the previous cases, we obtain the 
initial state of the system in the miscible phase, and then, imprint a singly
charged vortex at the center of the condensate of species $|1\rangle$. 
In experiments the generation of a vortex in either the condensate of 
Rb hyperfine TBEC was demonstrated by M. R. Matthews et al.~\cite{mat_99}. The 
initial density profiles of the condensates are as shown in 
Fig.~\ref{wvrtx_q1_cp1}(a) and (f). As to be expected the vortex is core-less, 
that is, condensate of $|2\rangle$ occupies the core of the vortex. Now, 
to observe the miscible-immiscible transition we quench $a_{12}$, and the 
density profiles during the quench are shown in Fig.~\ref{wvrtx_q1_cp1}.
As the value of $a_{12}$ is increased, the core size of the vortex increases,
and hence, larger number of atoms of species $|2\rangle$ occupy the vortex 
core. Since, the vortex is imprinted with the shell-condensate, as shown in 
Fig.~\ref{mis_immis}(c) and (d), there is an enhancement in the 
miscibility-immiscibility transition due to the centrifugal force associated 
with the vortex induced superflow. The enhancement is evident from the trend
in $\Lambda$ as shown in Fig.~\ref{lam_vs_time_q1}, and the effect is more
pronounced compared to the presence of vortices in both the species.
\begin{figure}[ht]
    \includegraphics[height=6.0cm]{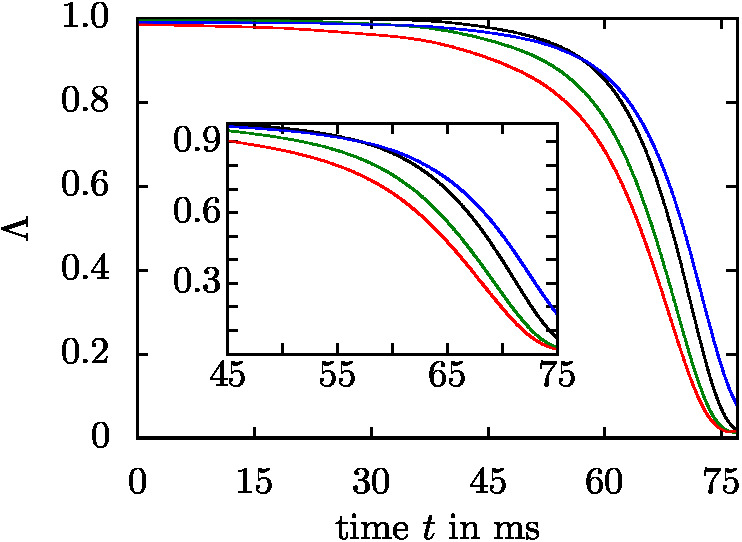}
    \caption{(Color online)
            Shows the variation of the overlap measure $\Lambda$ with time
            as $a_{12}$ of the Rb-Cs TBEC is quenched from $50a_{0}$ to 
            $175a_{0}$. The black curve shows this variation in absence of 
            defects in the condensates, the green curve shows it when a singly
            charged vortex is present at the center of both condensates; 
            whereas, the red curve shows the variation when the vortex is 
            present only in the Rb-condensate (shell-condensate) but not in the
            Cs-condensate (core-condensate), and the blue curve  shows the
            variation when the vortex is present only in the Cs-condensate 
            but not in the Rb-condensate.
            }
 \label{lam_vs_time_q1_rb_cs}
\end{figure}

 Another important observation is, vortex with higher charge leads to larger
enhancement in the miscible-immiscible transition. This is to expected since,
as discussed earlier, the centrifugal force is proportional to $l^2$, where
$l$ is the charge of the vortex. In the present case, there is an important 
observation, the vortices of higher charges are stable through the quench, 
and significant later times as well. This is in contrast to the case of single 
species condensates, where vortices of higher charges are dynamically unstable 
and decays in singly charged vortices with short time scales. The stability of 
a higher charge vortex in TBEC may be attributed to the immiscibility of the 
TBEC. Because, if the vortex decays to multiple vortices of lower charges it 
would increase the inter-species interaction energy due to the filling of 
the vortex cores. In short, TBEC supports higher charge vortex in the 
immiscible phase when the vortex is present in the shell-condensate.

 Similarly, for the Rb-Cs TBEC, we again obtain the initial equilibrium solution 
in the miscible phase, and a singly charged vortex imprinted
at the center of the Rb-condensate. It is to be mentioned here that, the Rb 
despite of having smaller atomic scattering is the shell-condensate due to the
smaller mass. In this case, the quench of $a_{12}$ leads to qualitatively 
similar results  as in Rb-hyperfine TBEC. That is, the core
size of the vortex increases during the quench, and the vortex induced 
superflow in the Rb-condensate enhances the phase-separation. This evident from
the trends in the values of $\Lambda$ shown in Fig.~\ref{lam_vs_time_q1_rb_cs}. 
\begin{figure}[ht]
    \includegraphics[height=4.0cm]{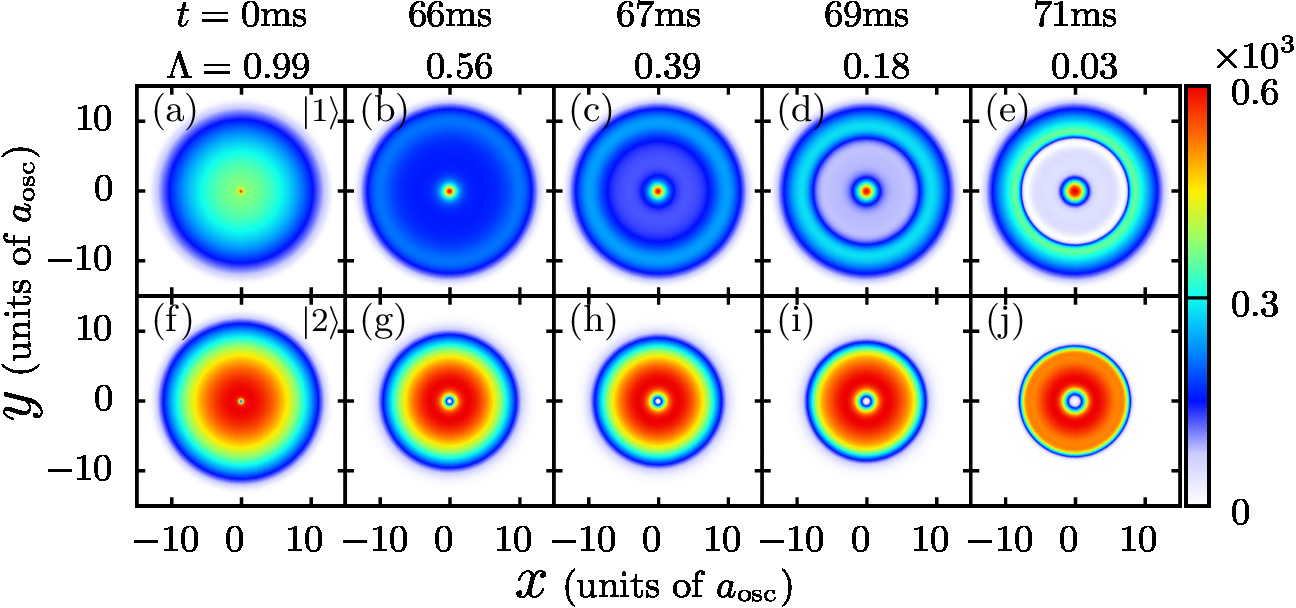}
    \caption{(Color online)
            Shows the time evolution of density profiles of the Rb-condensates
            in $|1\rangle$ and $|2\rangle$ states, during the quench of
            $a_{12}$, in presence of a singly charged vortex at the center of
            the core-condensate. The first and second row show the density
            profiles of the condensates in $|1\rangle$ and $|2\rangle$ states
            respectively, at $0$ms, $66$ms, $67$ms $69$ms and $71$ms. At these
            instants, the values of $a_{12}$ are $70a_{0}$, $96.7a_{0}$,
            $97.3a_{0}$ $98.0a_{0}$ and $98.7a_{0}$ respectively. Values of
            the overlap measure $\Lambda$ at corresponding time instants are
            mentioned at the top of each column. The color bars represent
            number density of atoms in the condensates in units of
            $a_{\rm osc}^{-2}$ where $a_{\rm osc} = 1.9\mu$m. 
            }
 \label{wvrtx_q1_cp2}
\end{figure}

 
\subsubsection{Vortex in core-condensate}

In this section we examine the dynamics of phase-separation when a vortex
is present in the core-condensate. For this, like in the previous case, the
initial state of the Rb-hyperfine TBEC is in the miscible phase, and a
singly charged vortex is imprinted at the center of the $|2\rangle$ condensate. 
The initial density profiles of the condensates are as shown in 
Fig.~\ref{wvrtx_q1_cp2}(a) and (f). We then quench the system by increasing
$a_{12}$ to drive the system to immiscible phase. The density profiles of the
condensates at different times during the quench are shown in 
Fig.~\ref{wvrtx_q1_cp2}. During this evolution, the core size of the vortex
increases, and an increasing number of atoms from species $|1\rangle$ occupy the
vortex core. Thus, in the immiscible phase of the TBEC, the density profile of
the condensate of the species $|1\rangle$ acquires a {\em bull's eye} structure 
as shown in Fig.~\ref{wvrtx_q1_cp2}(e). From the trend in $\Lambda$, shown in 
Fig.~\ref{lam_vs_time_q1}, it is evident that there is a suppression in 
phase-separation of the TBEC as the decrease in $\Lambda$ slower than the 
previous cases. The radially outward centrifugal force arising from the vortex 
leads to a larger radial size of the $|2\rangle$ condensate, and thus the atoms 
of $|1\rangle$ require larger inter-species repulsion energy to be the shell-
condensate at phase separation. In other words, the vortex induced superflow in 
the core-condensate is responsible for suppression of phase separation. 
>From similar computations, we also find the same trend in the Rb-Cs TBEC.
In fact, the effect of suppression is more pronounced in this system, this is
discernible by comparing the trends in the values of  $\Lambda$ plotted in 
Fig.~\ref{lam_vs_time_q1} and ~\ref{lam_vs_time_q1_rb_cs}.
 \begin{figure}[ht]
    \includegraphics[height=4.0cm]{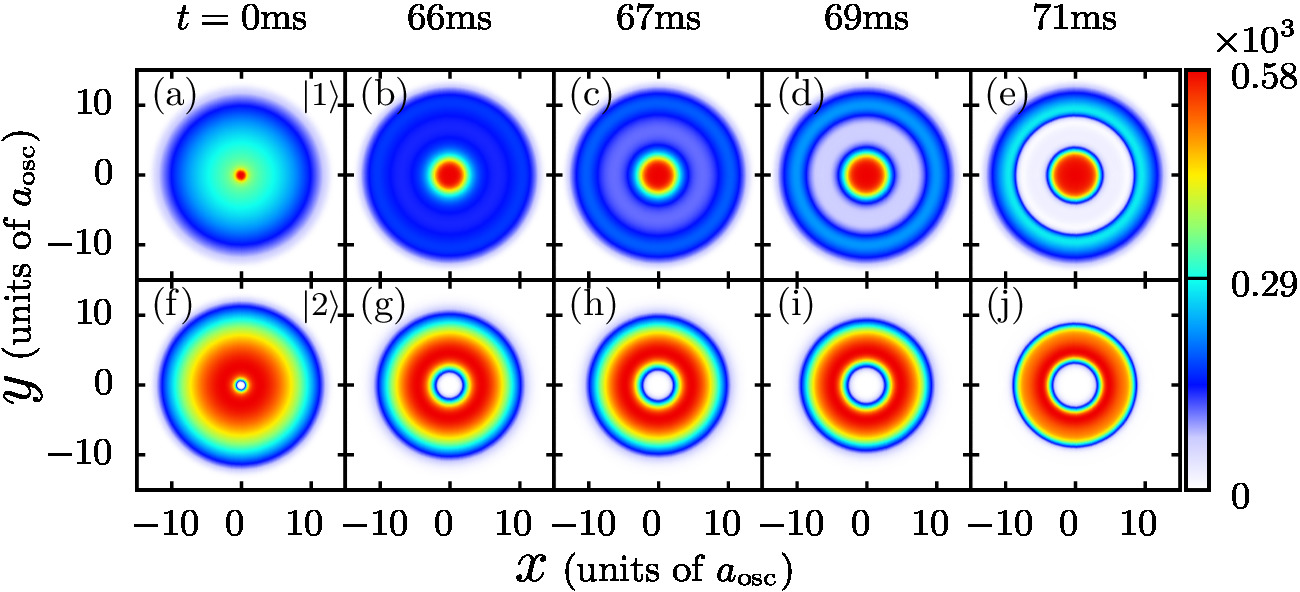}
    \caption{(Color online) 
            Shows time evolution of the density profiles of the TBEC in
            Rb-hyperfine states during the quench of $a_{12}$ in presence of a
            quadruply charged vortex at the center of the condensate of species
            $|2\rangle$. The color bars represent number density of atoms in 
            the condensate in units of $a_{\rm osc}^{-2}$ where 
            $a_{\rm osc} = 1.9\mu$m.}
 \label{wvrtx_q4_cp2}
 \end{figure}


\subsection{Higher charge vortex}

We now examine the dynamics of phase separation in presence of a
higher charge vortex, in particular with core-condensates. In 
experiments, doubly and quadruply charged vortices are generated using 
topological phase-imprinting technique~\cite{lea_02, kaw_04}.
The cases of vortices in both the condensates or only with shell condensate
are qualitatively similar to the cases of singly charged vortices. Like in 
the previous cases, we obtain initial equilibrium state of the
Rb-hyperfine TBEC in the miscible phase, but with a
quadruply charged vortex imprinted at the center of species $|2\rangle$ or the
core condensate. Then, we quench $a_{12}$ from $70a_{0}$ to $100a_{0}$
for miscibility-immiscibility phase transition in the TBEC. During the quench 
the core size of the vortex increases, and it gets filled with the atoms of
species $|1\rangle$ as shown in Fig.~\ref{wvrtx_q4_cp2}.
Hence, in the immiscible phase, the density profile of the condensate of 
species $|1\rangle$ has bull's eye structure with a higher density
core-region and a lower density ring outside the condensate of species 
$|2\rangle$. So, most of the atoms of species $|1\rangle$ occupy the core
region of the vortex, and is the consequence of larger core-size associated
with the higher charged vortex. Thus, the overall configuration has 
the density profile of the species $|2\rangle$ resembling the geometry
shell-condensate. In other words, the presence of quadruply charged vortex 
forces the species with lower intra-species interaction to occupy the edges, 
and the species with higher intra-species interaction to occupy the core region
by filling the vortex core. This can be referred to as the vortex induced 
partial position reversal at phase-separation. There is complete position
reversal when we consider a vortex with charge higher than $l=4$.
 \begin{figure}[ht]
 \includegraphics[height=4.0cm]{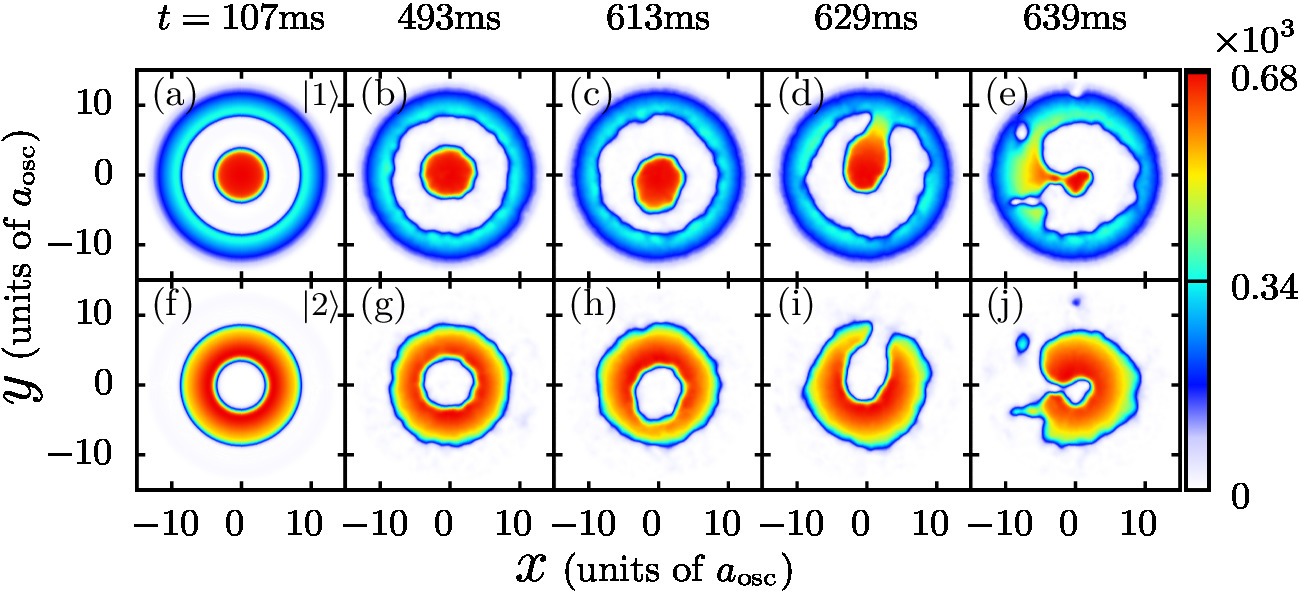}
    \caption{(Color online) 
            Shows post quench free dynamics of the phase-separated TBEC in 
            Rb-hyperfine states in presence of a quadruply charged vortex at 
            the center of the condensate of species $|2\rangle$. The color bars
            represent number density of atoms in the condensate in units of 
            $a_{\rm osc}^{-2}$, where $a_{\rm osc} = 1.9\mu$m.
            }
 \label{wvrtx_q4_cp2_evo}
 \end{figure}

The position reversed geometry is important to study as it provides a 
framework to investigate the dynamics related to Rayleigh-Taylor 
instability (RTI) in TBECs~\cite{gau_10, sas_09}. The instability sets in 
as the species initially occupying the core region is driven to the edge in the 
presence of the higher charged vortex when $a_{12}$ is quenched. In the case of 
quadruply charged vortex RTI is not observed, and the vortex induced azimuthal 
super-flow in the species $|2\rangle$ is responsible for inhibition of RTI at 
the interface of the condensates. This follows from the general result of 
suppression of RTI by the pressure gradient in the radial 
direction~\cite{cha_bk_13}, in the present case arising from the Coriolis force 
acting on atoms of species $|2\rangle$. In a related work, the suppression of 
RTI at the interface of rotating, immiscible, invisid classical fluids has been reported~\cite{tao_13}. Although RTI doesn't occur, the system exhibits a rich
dynamics associated with the prcession motion of the vortex during the post
quench free evolution of the TBEC as shown in Fig.~\ref{wvrtx_q4_cp2_evo}.
The condensates continue to be in the bull's eye and shell geometry 
respectively for sufficiently long time till $\approx 500$ms. However,
at later times, there is instability at the interface arising from the shear 
at the interface due to the superflow and decay of the higher charged vortex.
The density profiles of the TBEC at selected times during this later
evolution are evident from the density plots in Fig.~\ref{wvrtx_q4_cp2_evo}. 
We obtain qualitatively similar results for Rb-Cs TBEC in the presence of 
quadruply charged vortex in the Cs-condensate. 
Hence, in this case, as the TBEC is quenched to immiscible phase, the 
Rb-condensate takes the bull's eye structure, and the density profile of the 
Cs-condensate resembles with shell-condensate.


\section{Conclusions}\label{Conclusions}
In presence of singly charged vortex in the shell-condensates of the TBECs, 
the centrifugal force associated with the vortex induced azimuthal superflow 
enhances the miscibility-immiscibility phase-separation. The same force 
resulting from the vortex induced superflow in the core-condensates suppresses 
the phase-separation. However, there is a net enhancement when singly charged
vortices are present in both of the species. Compared to the Rb hyperfine TBEC, 
in Rb-Cs TBEC the centrifugal force experienced by Rb atoms is stronger. 
Hence, the enhancement or suppression of phase-separation due to the presence 
of vortex is more prominent in Rb-Cs TBEC. The quadratic dependence of the 
centrifugal force on the vortex charge, ensures the obtained results are 
independent on the sense of circulation of the super-flows. The results 
from the Rb-Cs TBEC are generic to the TBECs in which the species have 
considerable mass difference, and different intra-species interactions. 
Similarly, the results of the Rb hyperfine TBEC is generic to other TBEC of
two hyperfine states, isotopes of the same elements or different atoms with 
nearly equal mass and scattering length. Thus, the cases considered is 
representative of other TBEC. In presence of a vortex of quadruply charged
vortex in the core-condensate, a phase-separated state of the TBECs is 
obtained in which the components of the TBECs partially swap their positions 
in the shell-structured geometry in comparison with the case when the vortex 
is absent. From the post quench free dynamics, at later times there is an 
instability at the interface and decay of the quadruply charged vortex.

\begin{acknowledgments}

We thank S. Pal, K. Suthar and R. Bai for useful discussions. The results
presented in this paper are based on the computations using Vikram-100, the
100TFLOP HPC Cluster at Physical Research Laboratory, Ahmedabad, India.

\end{acknowledgments}

\bibliography{ph_sep}{}
\bibliographystyle{apsrev4-1}

\end{document}